# Gate-tunable, normally-on to normally-off memristance transition in patterned LaAlO$_3$/SrTiO$_3$ interfaces

P. Maier,[1] F. Hartmann,[1,a)] J. Gabel,[2] M. Frank,[1] S. Kuhn,[1] P. Scheiderer,[2] B. Leikert,[2] M. Sing,[2] L. Worschech,[1] R. Claessen[2] and S. Höfling[1,3]

[1]*Technische Physik, Physikalisches Institut and Röntgen Center for Complex Material Systems (RCCM), Universität Würzburg, Am Hubland, D-97074 Würzburg, Germany.*

[2]*Experimentelle Physik 4, Physikalisches Institut and Röntgen Center for Complex Material Systems (RCCM), Universität Würzburg, Am Hubland, D-97074 Würzburg, Germany.*

[3]*SUPA, School of Physics and Astronomy, University of St. Andrews, St. Andrews, KY16 9SS, United Kingdom.*

We report gate-tunable memristive switching in patterned LaAlO$_3$/SrTiO$_3$ interfaces at cryogenic temperatures. The application of voltages in the order of a few volts to the back gate of the device allows controlling and switching on and -off the inherent memory functionality (memristance). For large and small gate voltages a simple non-linear resistance characteristic is observed while a pinched hysteresis loop and memristive switching occurs in an intermediate voltage range. The memristance is further controlled by the density of oxygen vacancies, which is tuned by annealing the sample at 300 °C in nitrogen atmosphere. Depending on the annealing time the memristance at zero gate voltage can be switched on and off leading to normally-on and normally-off memristors. The presented device offers reversible and irreversible control of memristive characteristics by gate voltages and annealing, respectively, which may allow to compensate fabrication variabilities of memristors that complicate the realization of large memristor-based neural networks.

---

a) Corresponding author; electronic mail: fabian.hartmann@physik.uni-wuerzburg.de.

The interplay of charge, spin and orbital degrees of freedom in transition metal oxides (TMOs) leads to a variety of physical properties including the formation of quasi two-dimensional electron systems (q2-DES),[1,2] superconductivity,[3,4] ferroelectricity[5,6] and magnetism.[7-9] The formation of a q2-DES was first observed at the $LaAlO_3/SrTiO_3$ (LAO/STO) interface by Ohtomo and Hwang in 2004.[1] Electronic reconstruction due to the polar discontinuity between the LAO film and the STO substrate has been suggested as an explanation for the formation of the metallic interface state.[10] However, depending on deposition parameters, oxygen vacancies in the STO substrate may also provide charge carriers and contribute to the q2-DES formation.[7,11-13] Controlling the drift of the oxygen vacancies with an electric field enables the realization of resistance switching, that can be connected to memristor theory.[14,15] Memristors are promising candidates to implement synaptic functionalities in artificial neural networks[16,17] and enable the implementation of brain-inspired recognition and classification tasks.[18] They are characterized by a state-dependent memory resistance (memristance) which is governed by the previous charge flow through the device and leads to a characteristic pinched hysteresis loop in the current-voltage-plane.[14,19] Up to now, memristors have been realized in various material systems including two-dimensional $MoS_2$,[20-22] ZnO nanorods,[23] organic materials[24] and transition metal oxides like $HfO_x$,[25] $Ta_2O_5$[26] or STO.[27-29] Resistive switching in transition metal oxides is explained by the drift of mobile anions and cations in an electric field[30] and enables the realization of logic gates,[31] memories[32] or in-memory-adders.[33] Memristive functionality has also been reported for planar LAO/STO interfaces with non-ohmic contacts.[34,35] For the scalability of memristive networks, the variability in the fabrication process is essential.[36] Device variability may be compensated by tuning characteristic voltages such as the set voltage with additional gates, which may be beneficial for fault tolerant architectures.[20]

The memristor proposed here is based on patterned LAO/STO interfaces and enables controlling of the memristance with back gate voltages of only a few volts. For large and low back gate voltages, the area of the pinched hysteresis loop tends to zero and the device shows non-linear resistances without inherent memory. Due to the gate-tunable memristance, the device is suitable as artificial synapse with learning processes supervised by a control voltage. In addition, the memory functionality is tuned by creating oxygen vacancies when the sample is annealed at 300 °C in nitrogen atmosphere, which allows implementing normally-on and normally-off memristors showing non-linear resistances and pinched hysteresis loops at zero gate voltage, respectively.

Fig. 1(a) shows a three-dimensional scheme of the device consisting of a conducting wire and two lateral gates at the LAO/STO interface. Patterning of the device is based on the technique presented by Schneider et al. in Ref. 37. Starting from a $TiO_2$-terminated STO-substrate, first two unit cells of LAO are grown by pulsed laser deposition (PLD) with a KrF eximer laser ($\lambda = 248$ nm) in $1*10^{-3}$ mbar oxygen partial pressure at a substrate temperature of 780 °C, a laser fluency of 1.0 $Jcm^{-2}$, a pulse repetition rate of 1 Hz and a target-substrate distance of 53 mm.[38] Single crystalline LAO is used as target material. Via electron beam lithography the device layout is patterned and after resist development, 11 nm of amorphous $SiO_2$ are deposited using an e-beam evaporation technique. Hence, the two unit cells of LAO are only covered by $SiO_2$ at well-defined positions. After annealing the sample for one hour at 780 °C to get rid of remnants from the photoresist treatment, the LAO growth via PLD is continued. Amorphous LAO is deposited on the areas covered by $SiO_2$ whereas four epitaxial unit cells of LAO are grown on the areas not covered by $SiO_2$. To ensure full oxidation, the sample is slowly cooled down to room temperature in 500 mbar $O_2$ with the temperature being held constant for one hour at an elevated temperature of 600 °C. By this technique the critical thickness of four unit cells of epitaxial LAO for the formation of the q2-DES[2] is only exceeded at specific areas and a conducting wire can be realized. If not stated differently, the present measurements were conducted at 4.2 K in the dark. The wire was contacted by depositing Ti/Au on top of the LAO. Fig. 1(b) shows an electron microscopy image of the resist on top of the two unit cells of LAO, prior to the $SiO_2$ and the second LAO layer deposition. The critical layer thickness of four unit cells of crystalline LAO, leading to the formation of the q2-DES, is later only exceeded in the orange and green shaded areas, corresponding to the wire and lateral gates, respectively. The lateral gates on the left and right side of the wire allow the realization of transistors with high efficiencies,[39] but were set floating for the present studies. The gate-dependent measurements were carried out with the back gate (see Fig. 1(a)) at the bottom of the STO substrate.

Figs. 1(c) and (d) show Hall mobilities and charge carrier densities of the as-grown sample (solid symbols) and after annealing the sample for 30 s at 360 °C in nitrogen atmosphere (open symbols). The annealing creates oxygen vacancies which increases the charge carrier density. The existence of oxygen vacancies after annealing likewise becomes apparent in the carrier freeze-out observed at low temperatures in particular in the annealed samples.[40] As the vacancies act as scattering centers, the mobilities of the annealed samples are as well reduced in the temperature range from 2 to 300 K.

A typical current-voltage-characteristics of the device is shown in Fig. 2(a). The voltage $V_b$ (voltage source) is applied to one terminal of the wire and the other terminal is connected to ground. Performing a closed voltage sweep cycle, a pinched hysteresis loop in the *I-V*-plane as characteristic of memristors is evident.[19] The pinched hysteresis loop can be explained by charging and discharging of interfacial traps (e.g. metal-induced gap states, oxygen vacancies in the LAO or STO) in the contact region.[34] In analogy to Ref. 34, the device can be represented by an equivalent circuit consisting of the wire-resistance $R_{\text{wire}}$ and two non-ohmic resistances $R_{\text{LAO,1}}$ and $R_{\text{LAO,2}}$ that represent Schottky-barriers at the metal/LAO/q2-DES interface (see inset of Fig. 2(a)). These Schottky-barriers host interfacial traps which become charged via tunneling at negative voltages leading to larger Schottky barriers and contact resistances $R_{\text{LAO,1}}$ and $R_{\text{LAO,2}}$. Consequently, the total resistance of the equivalent circuit increases as

$$R_{tot} = R_{wire} + R_{LAO,1} + R_{LAO,2}. \tag{1}$$

Discharging of the interfacial traps at positive voltages on the other hand reduces the total resistance. The resistance variation during the voltage sweep cycle leads to the observed pinched hysteresis loop in the current-voltage-characteristics. Resistive switching based on the variation of Schottky barrier heights and widths was also reported for STO-based memristive devices, where changes of spatial donor concentrations[29] or the charging of trap states near the interface between the STO and the electrode[41] lead to the observation of different resistive states.

Increasing driving frequencies of the voltage sweep reduce the area of the hysteresis as displayed in Fig. 2(b). For larger frequencies, the charge in the interfacial traps remains constant due to the low tunneling rate compared to the driving frequency. Consequently, the total resistance of the equivalent circuit remains constant during one sweep cycle and the area of the hysteresis is almost reduced to zero for a frequency of 100 Hz. Hence the switching time of the presented memristor is in the order of 10 ms, which is larger than the switching time of 100 µs in unpatterned LAO/STO interfaces with Pt-contacts.[35] The lowest reported switching time is 10 ns in $TaO_x$-based memristors.[26] Smaller switching times of the present device may be realized by increasing the operation voltage range.

In contrast to previous reports on resistance switching based on non-ohmic contacts of a planar q2-DES,[34,35] we report resistance switching of a patterned q2-DES with memristance control by the application of back gate voltages in the range of a few volts only. Realizing a wire with lateral width of 200 nm enables efficient resistance modulation of the wire with relatively low voltages applied to the back gate due to a geometry induced increase of the capacitance between the wire and the back gate.[42] Fig. 3(a) shows current-voltage-characteristics for different back gate voltages, obtained after annealing the device for six minutes in nitrogen atmosphere at 300 °C. Varying areas of the pinched hysteresis loop are observed when tuning the back gate voltage. Fig. 3(b) shows the areas of the pinched hysteresis loops versus the gate voltage (orange squares correspond to the annealing time of 6 min and the *I-V*-curves presented in Fig. 3(a)). For a gate voltage of -3 V the area is maximal. The dependency of the area on the gate voltage is explained by resistance changes of the wire ($R_{wire}$ in Eq. (1)). Below -3 V, the wire becomes depleted which increases the wire resistance ($R_{wire}$(0 V) < $R_{wire}$(-3 V)). At constant bias voltage the voltage drop across the Schottky-barriers $R_{LAO,1}$ and $R_{LAO,2}$ consequently decreases and charging of the interfacial traps becomes less effective. The memory effect vanishes and the area of the hysteresis tends to zero. For larger gate voltages (i.e. above -3 V), charge carriers are accumulated in the wire leading to a Schottky-barrier height reduction at the metal/LAO/q2-DES interface, hence the contacts become more and more ohmic. The memory effect vanishes and the area of the hysteresis tends to zero again. Controlling the resistance of the wire with the gate voltage allows to switch the memristance on and off in a reversible manner.

To further investigate fabrication-dependent memristance-control, the device was annealed at 300 °C in nitrogen atmosphere. Fig. 3(c) displays *I-V*-curves for a gate voltage of zero and different annealing times. The area of the pinched hysteresis loop is maximal for a time of 11 min. Depending on the annealing time, the maxima occur at different gate voltages (see Fig. 3(b)). For longer annealing times the maximal area is observed at higher back gate voltages The shifting of the maximal area of the pinched hysteresis loop towards larger voltages can be explained by the generation of oxygen vacancies during the annealing of the sample (see Figs. 1(c) and (d)) which release mobile electrons that screen the electric field of the back gate. Fig. 3(d) shows current-voltage-characteristics for the combinations of annealing times and gate voltages with maximal area in Fig. 3(b). Varying the gate voltage allows to observe similar *I-V*-curves for different annealing times. The enhanced screening by mobile electrons can thus indeed be compensated with larger back gate voltages. Interpolating the data in Fig. 3(b) allows to determine annealing times to realize normally-off and normally-on memristors that show zero and non-zero areas for a gate voltage of zero, respectively. For an annealing time of 9 min, the maximum area is expected to be at zero back gate voltage. Annealing times larger than 25 min should allow realizing normally-off memristors, which only show pinched hysteresis loops in the current-voltage-characteristics when applying positive back gate voltages.

With their state-dependent resistance, memristors enable emulating synaptic functionalities which may be implemented in novel, brain-inspired computing architectures.[16,18] To investigate the impact of gate-tunable memristors on synaptic functionalities, the device is excited with voltage pulses while applying constant back gate voltages. Fig. 4(a) shows the conductance of the device versus pulse number when applying voltage pulses with width and amplitudes of 25 µs and 8 V, respectively. Initially ($N = 0$), the conductance is low and increases for larger $N$. The conductance of the device can be increased by voltage pulses, which is essential to implement synaptic functionalities. In addition, the gate voltage allows to control the conductance for $N = 50$ (see Fig. 4(b)). For larger gate voltages the conductance increases.

Applying voltage pulses with negative amplitudes of -6 V and widths of 250 µs enables to successively reduce an initially large conductance as depicted in Fig. 4(c). Here, the gate voltage allows to tune the conductance reduction per pulse. Fig. 4(d) shows the time constants of the conductance reduction that were obtained by fitting the data in Fig. 4(c) with exponential functions. For increasing gate voltages the time constant is lowered. The required number of pulses to induce depression (reducing synaptic strength) of the artificial synapse depends sensitively on the gate voltage which allows implementing synapses that store information very effectively (negative gate voltages) or synapses that are depressed after a few pulses (positive gate voltages).

Controlling the conductance of the device with voltage pulses enables emulating synaptic plasticity which is key for learning and memory in neural networks.[43-45] The dependency of the memristance change on the back gate voltage is beneficial to tune learning processes of different synapses. As an example, negative back gate voltages allow to suppress the conductance increase in Fig. 4(b) and require the excitation with many pulses to reduce the conductance (see Fig. 4(d)). Memristors with additional gates have not been studied intensively in the literature, but control of the set-voltage was reported in Ref. 20. The presented memristance control is due to accumulation and depletion of a narrow wire and may be transferrable to other memristor realizations that show synaptic functionalities at room temperature.

In summary, we present a memristor based on a patterned LAO/STO heterostructure that allows controlling the memristance reversibly by back gate voltages and irreversibly by annealing the device in nitrogen atmosphere. Tuning the area of the pinched hysteresis loop in the current voltage characteristics leads to normally-on and normally-off memristors which show non-zero and zero areas at zero gate voltage, respectively. The control of memristor-characteristics may be beneficial to compensate device variabilities that are induced by fabrication processes and impede large scalability of memristor networks. Direct application of the device is hindered by the low operation temperature, which may be overcome by implementing Pt contacts on top of the LAO layer and thus increasing the Schottky-barriers. The gate-tunability should be directly transferrable to other memristor realizations with top gate geometry.

**Acknowledgements**

The authors gratefully acknowledge the support from the state of Bavaria as well as from the Deutsche Forschungsgemeinschaft (FOR1162 and SFB1170).

**Figures and figure captions:**

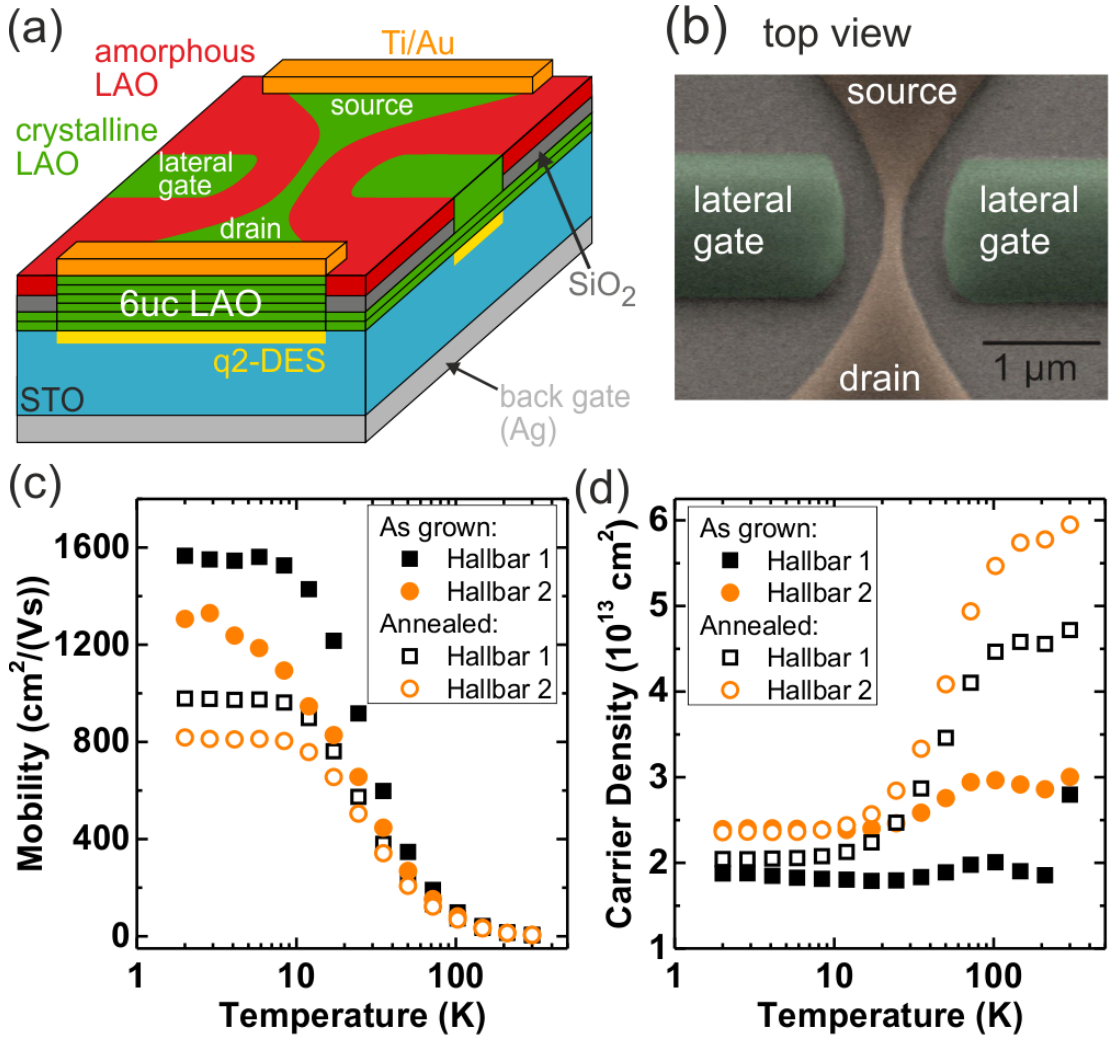

Fig. 1. (a) Sketch of the device depicting the wire, lateral gates, Ti/Au contacts and back gate. The conducting wire is formed at sites with 6 unit cells of crystalline LaAlO$_3$ (green) on SrTiO$_3$. Other areas are covered with amorphous SiO$_2$ (dark gray) and amorphous LaAlO$_3$ (red). (b) Electron microscopy image of the device with the wire (orange) and two lateral gates (green). (c) Hall mobilities of the as-grown sample (solid symbols) and after annealing the sample for 30 s at 360 °C in nitrogen atmosphere (open symbols). (d) Corresponding charge carrier densities to panel c.

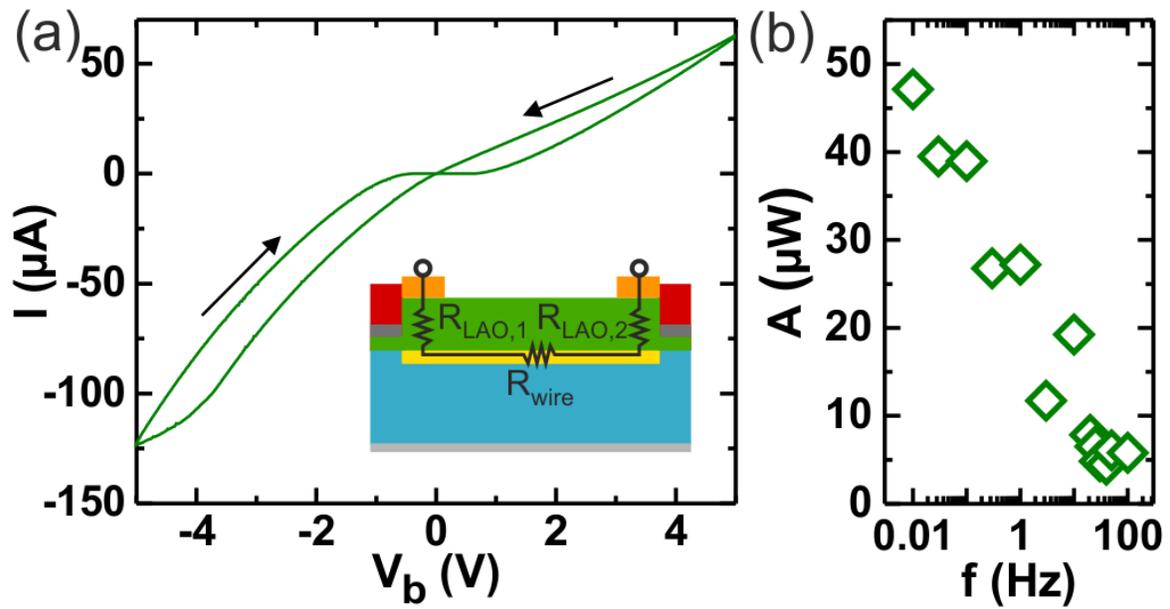

Fig. 2. (a) Current-voltage-characteristics of the device. The arrows indicate the sweep cycle direction. A pinched hysteresis loop is evident. Inset: Equivalent circuit of the device consisting of two contact resistances ($R_{LAO,1}$ and $R_{LAO,2}$) and the gate-tunable wire resistance ($R_{wire}$). (b) Area of the pinched hysteresis loop versus the voltage sweep frequency. The area vanishes for large driving frequencies.

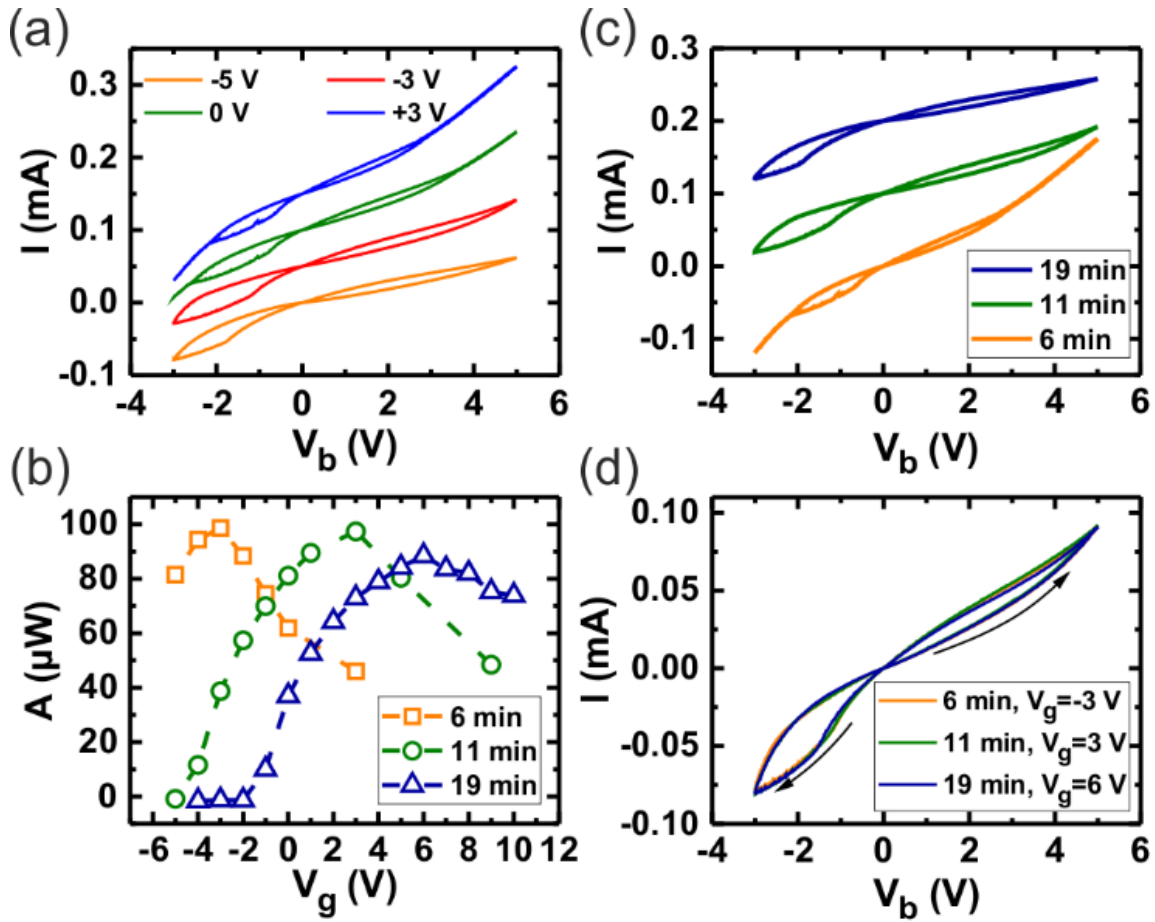

Fig. 3. (a) Current-voltage-characteristics for different back gate voltages obtained after an annealing time of 6 minutes at 300 °C in a nitrogen atmosphere. The area of the pinched hysteresis loop varies with the gate voltage. The curves are offset by 0.05 mA. (b) Area versus gate voltage for different annealing times. The area is maximal at back gate voltages -3, 3 and 6 V for annealing times 6, 11 and 19 min, respectively. For increasing annealing time, the maximal area is observed at larger gate voltages. (c) Current-voltage-characteristics for different annealing times and zero gate voltage applied. The curves are offset by 0.1 mA for clarity. (d) Current-voltage-characteristics for the combinations of annealing time and gate voltage that show maximal areas in panel b.

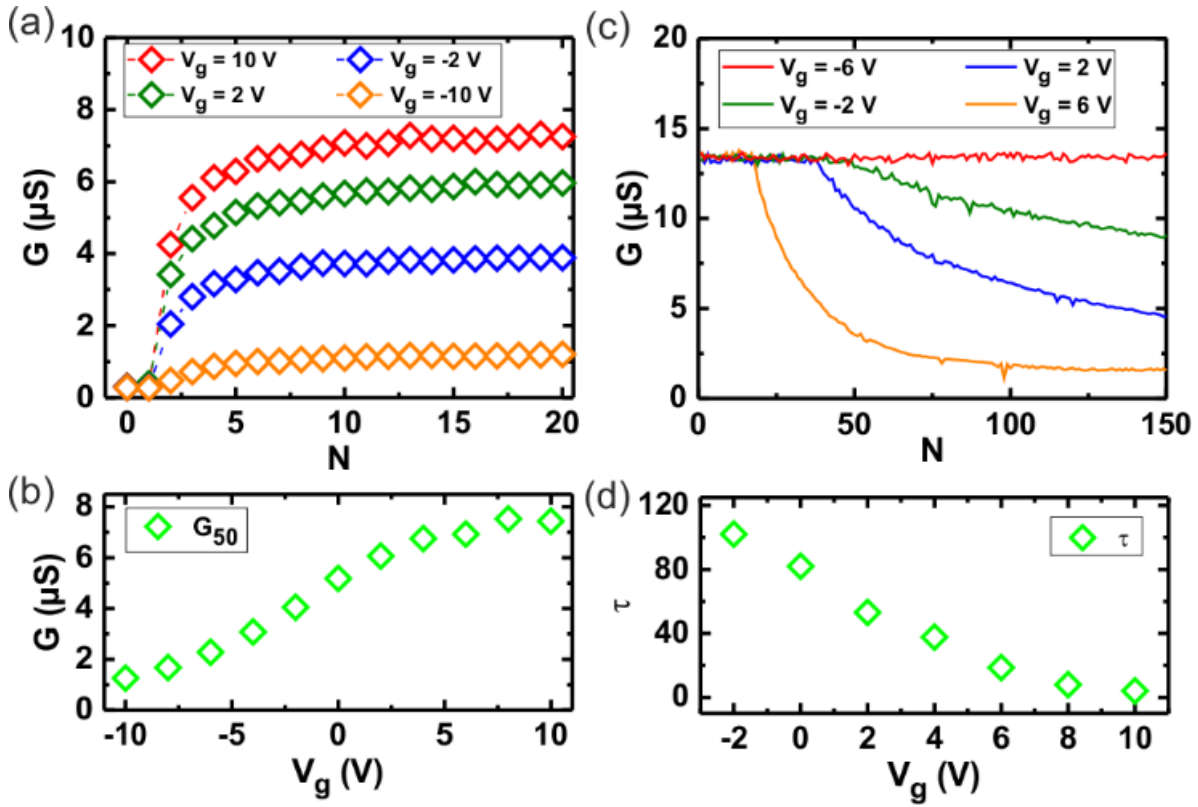

Fig. 4. (a) Conductance *G* versus pulse number for positive voltage pulses and different back gate voltages. Larger gate voltages lead to enhanced conductance values after applying 50 pulses as shown in (b). (c) Conductance versus pulse number for negative voltage pulses and different gate voltages. Tuning the gate voltage allows to control the time constant $\tau$ of the conductance reduction. (d) Time constant $\tau$ of the data in c versus gate voltage. Large gate voltages reduce the time constant.